\documentclass[aps,pra,superscriptaddress,showpacs,floatfix,longbibliography,notitlepage,twocolumn]{revtex4-2}
\usepackage{dsfont}
\usepackage{cprotect}
\usepackage{graphicx}
\usepackage{amsmath,mathtools}
\usepackage{commath}
\usepackage{comment} 
\usepackage{amssymb}
\usepackage{bm}
\usepackage{algorithm}
\usepackage{algpseudocode}
\newcommand{\e}{\text{e}}
\newcommand{\beq}{\begin{equation}}
\newcommand{\eeq}{\end{equation}}
\newcommand{\beqnn}{\begin{equation*}}
\newcommand{\eeqnn}{\end{equation*}}
\newcommand{\bea}{\begin{eqnarray}}
\newcommand{\eea}{\end{eqnarray}}
\newcommand{\nn}{\nonumber}
\newcommand{\beann}{\begin{eqnarray*}}
\newcommand{\eeann}{\end{eqnarray*}}
\newcommand{\bes} {\begin{subequations}}
\newcommand{\ees} {\end{subequations}}

\newcommand{\bra}[1]{\langle #1 |}
\newcommand{\ket}[1]{| #1 \rangle}

\newcommand\sjl{{\bf j} \in \Omega_{(\!K,q\!)}} 

\usepackage{xcolor}

\newcommand{\ih}[1]{{\color{black}#1}}

\usepackage{qcircuit}

\begin{document}
\title{A simple quantum simulation algorithm with near-optimal precision scaling}

\author{Amir Kalev}
\affiliation{Information Sciences Institute, University of Southern California, Arlington, VA 22203, USA}
\author{Itay Hen}
\email{itayhen@isi.edu}
\affiliation{Information Sciences Institute, University of Southern California, Marina del Rey, California 90292, USA}
\affiliation{Department of Physics and Astronomy and Center for Quantum Information Science \& Technology, University of Southern California, Los Angeles, California 90089, USA}

\begin{abstract}
\noindent Quantum simulation is a foundational application for quantum computers, projected to offer insights into complex quantum systems beyond the reach of classical computation. However, with the exception of 
Trotter-based methods, which suffer from suboptimal scaling with respect to simulation precision, existing simulation techniques are, for the most part, too intricate to implement on early fault-tolerant quantum hardware. We propose a quantum Hamiltonian dynamics simulation algorithm that aims to  be both straightforward to implement and, at the same time, have near-optimal scaling in simulation precision.
\end{abstract}

\maketitle

\section{Introduction}
Quantum computers are widely believed to have unique advantages over classical computers when it comes to simulating Hamiltonian dynamics, due to their inherent quantum nature, which allows them to efficiently represent complex quantum states and the evolution thereof. Such simulations hold the potential to solve intricate problems in quantum chemistry, materials science, and physics, making quantum computers potentially very powerful future tools for advancing scientific knowledge and tackling challenges beyond the capabilities of classical computing.

The precise manner in which Hamiltonian dynamics can be approximated on quantum devices varies considerably between approaches, as each is tailored to different resource constraints, runtime requirements, and precision needs. Perhaps the most straightforward and resource-efficient technique is that of Trotterization, in which case the time-evolution operator is approximated using Trotter-Suzuki product formulas~\cite{Lloyd:96,PhysRevX.11.011020}. This approach is simple to implement; however, it suffers from a suboptimal scaling of the resources required for its execution with error tolerance. A more advanced and vastly popular technique is quantum signal processing (QSP)~\cite{QSP}, which leverages a sequence of unitary operations combined with classical preprocessing to directly approximate the exponential of a Hamiltonian.  
Complementing QSP, the linear combination of unitaries (LCU) framework~\cite{Berry1} encompasses some of the most powerful quantum simulation methods~\cite{Berry1,Kalev2021quantumalgorithm,PRXQuantum.2.030342,PhysRevLett.131.150603,Chakraborty2024implementingany}. In LCU-based approaches, the time evolution operator is expressed as a linear combination of efficiently implementable unitary operators, with ancillary qubits facilitating the simulation through oblivious amplitude amplification~\cite{OAA}. This method is particularly advantageous for simulating Hamiltonians in quantum chemistry and materials science, where the decomposition of the Hamiltonian into simpler components is feasible.

Despite enjoying near-optimal scaling of resources with respect to target precision --- typically requiring only a logarithmic number of ancillary qubits --- these more advanced approaches present significant challenges for implementation on near-term quantum computing devices. For example, in the typical LCU-based approach, the `select' unitary transformations are (multi-qubit) control of  Pauli strings. These control operations, which involve different Pauli operations acting on different qubits, translate to a non-trivial overhead in resources, which then hinders the technique's otherwise straightforward applicability on near-term quantum computing devices. 

In this work, we present a novel LCU-based quantum simulation algorithm designed to be both straightforward to implement, by generally requiring only control-not (CNOT) operations as the select unitaries, while at the same time offering the benefits of near-optimal resource scaling, making dynamical simulations potentially more accessible for near-term devices. 

The simulation algorithm we present here builds on two recent studies~\cite{Kalev2021quantumalgorithm,PRXQuantum.2.030342} that have shown how the expansion of the unitary time-evolution operator in powers of its off-diagonal strength~\cite{ODE,ODE2,pmr} combined with the use of the concept of divided differences~\cite{dd:67,deboor:05} leads to resource-efficient quantum simulation algorithms whose complexity is on par and, in some cases, superior to other state-of-the-art optimal-precision simulation protocols~\cite{Berry1,2018arXiv180500675H}. The approach we present here, which we refer to as the permutation matrix representation (PMR) simulation method, similarly utilizes a series expansion of the quantum time-evolution operator in its off-diagonal elements. However, in contrast to the work depicted in Refs.~\cite{Kalev2021quantumalgorithm,PRXQuantum.2.030342}, here we utilize a powerful approximation of divided differences which allows us to render those in terms of sums of phases, which in turn leads to simplifications in the select unitary operations. 

As we demonstrate, when the above approximation is combined with the LCU technique, the resulting simulation algorithm enjoys both near-optimal scaling in terms of the overall simulation error and simple-to-implement CNOTs interleaved with controlled phases as the select unitary operations. 

The paper is organized as follows. In Sec.~\ref{sec:pmr}, we provide an overview of the permutation matrix representation technique, which serves as a foundation for our algorithm. In Sec.~\ref{sec:algo}, we present the Hamiltonian dynamics algorithm, which we construct using PMR, and discuss the divided difference calculation including cost analysis and various extensions. In Sec.~\ref{sec:results} we demonstrate the scaling advantages of our method by analyzing in detail the algorithm as it is applied to two specific models, namely the Rydberg atom Hamiltonian and dipolar molecules in optical lattices. A discussion and some conclusions are given in Sec.~\ref{sec:summary}.

\section{Permutation matrix representation of the time evolution operator\label{sec:pmr}}
Before delving into the technical specifics of our algorithm, we first briefly present the key components of the PMR approach, which we use to expand matrix exponentials in a series~\cite{pmr,Kalev2021quantumalgorithm,PRXQuantum.2.030342}. For concreteness, we restrict our attention here to time-independent Hamiltonians and defer the discussion of how the method can be extended to the time-dependent case to the concluding section. 

The PMR expansion consists of several steps, the first of which is choosing a preferred basis, the `computational basis', in which the Hamiltonian matrix is represented, and casting the Hamiltonian $H$ as a sum of a diagonal operator and an off-diagonal operator in that basis. Hereafter, we denote the set of computational basis states as $\{\ket{z}\}$. Intuitively, the diagonal part of the Hamiltonian corresponds to a `classical' Hamiltonian, whereas its off-diagonal part governs the non-trivial `quantum' dynamics of the system in the chosen basis. The off-diagonal component is further decomposed to a sum of products of diagonal and permutation operators~\cite{pmr}. This defines the PMR form of the Hamiltonian:
\beq\label{eq:PMRofH}
H=D_0+\sum_{i=1}^M D_i P_i \,,
\eeq
where the $D_i$'s are diagonal operators and the $P_i$'s are permutation operators, i.e., operators that permute the basis states. For example, for qubit systems, if one chooses the computational basis to be the eigenbasis of the Pauli-$Z$ operators, then the  $P_i$ operators are strings of Pauli-$X$ operators. The decomposition Eq.~\eqref{eq:PMRofH} can be carried out efficiently for any physical Hamiltonian~\cite{pmr}. Next, consider the evolution of a quantum state under $H$ for duration $t$, which is then split to a sequence of $r$ repeated short-time evolution operators, each evolving the state by a short time period $\Delta t=t/r$, the value of which we determine later on in a manner designed to optimize resources (we discuss this in detail in Sec.~\ref{sec:cost}).

The PMR decomposition of the Hamiltonian allows us to write the (short) time evolution operator \hbox{$U(\Delta t)=e^{-i H\Delta t}$} in an off-diagonal series expansion:
\begin{align}
e^{-i H\Delta t} &= \sum_{n=0}^{\infty}\frac{(-i \Delta t)^n}{n!} H^n=\sum_{n=0}^{\infty}\frac{(-i \Delta t)^n}{n!}  \Big(\sum_{i=0}^M D_i P_i\Big)^n ,\end{align}
where in the last step we identify $P_0$ with the identity operator. After some algebra, this expansion (see Refs.~\cite{ODE,pmr} for a complete and detailed derivation) may be expressed as~\cite{Kalev2021quantumalgorithm} 
\begin{align}\label{eq:Uod}
U (\Delta t) =\sum_{q=0}^{\infty}  \frac{\Delta t^q}{q!}\sum_{{\bf i}_q}   \Gamma_{{\bf i}_q} P_{{\bf i}_q} A_{{\bf i}_q}.
\end{align}
Here, the boldfaced index ${\bf i}_q = (i_1,\ldots,i_q)$ is a tuple of indices $i_j$, with $j=1,\ldots, q$, each ranging from $1$ to $M$ and $P_{{\bf i}_q} := P_{i_q} \cdots P_{i_2}P_{i_1}$ is an ordered product of off-diagonal permutation operators. In addition, the operators $A_{{\bf i}_q}$ are the diagonal operators 
\beq\label{eq:Aiq}
A_{{\bf i}_q} =  \sum_z \frac{d_{{\bf i}_q}}{\Gamma_{{{\bf i}_q}}}  \frac{q!}{\Delta t^q} \e^{-i  \Delta t [E_{z_0}, E_{z_1},\ldots,E_{z_q}]} |z\rangle \langle z| \,,
\eeq
where $\e^{-i  \Delta t [E_{z_0},\ldots,E_{z_q}]}$ is the \emph{divided difference} of the exponential function~\cite{dd:67,deboor:05} $f(\cdot)=\e^{-i \Delta t (\cdot)}$ over the multi-set  $\{E_{z_0},\ldots, E_{z_q}\}$, where  $E_{z_j}=\langle z_j | D_0|z_j\rangle$ with \hbox{$\ket{z_j}=P_{i_j}\cdots P_{i_1}\ket{z}$}, and we used the convention that ${z_0}=z$. 
The reader is referred to Appendix~\ref{app:dd} for additional details on divided differences. Note that the $z_j$'s in $\ket{z_j}$ and in $E_{z_j}$ should actually be denoted by $z_{{\bf i}_j}(z)$, however for conciseness we are using the abbreviations $z_j$. Lastly, we have denoted the product of off-diagonal matrix elements as $d_{{\bf i}_q}=\prod_{j=1}^q d_{i_j}(z_j)$,  where 
\beq\label{eq:dj}
d_{i_j}(z_j) = \langle z_j | D_{i_j}|z_j\rangle
\eeq
can be considered  the `hopping strength' of $P_{i_j}$ with respect to $|z_j\rangle$, and we have defined the real-valued coefficients
$\Gamma_{{\bf i}_q} = \prod_j \Gamma_{i_j}$ where $\Gamma_i = \max_z |d_{i}(z)|$.  

In what follows, we use the expanded form of $U(\Delta t)$ given in Eq.~\eqref{eq:Uod} to show that the time evolution operator can be formulated as a linear combination of simple-to-execute unitary operators which in turn allows us to utilize the LCU technique to approximate the time-evolution operator.

The main technical contribution of this paper is devising an algorithm that implements the LCU with near-optimal scaling and at the same time makes use of only simple-to-implement unitary operators, avoiding complicated classical computations on quantum registers or involved select unitaries. We present our algorithm next.

\newpage
\section{The simulation algorithm\label{sec:algo}}

\ih{\subsection{Overview and summary of the algorithm} 

We begin by providing a general overview and summary of the main components of the simulation algorithm, which will be covered in detail in subsequent subsections. Given an input Hamiltonian in its PMR decomposition, Eq.~\eqref{eq:PMRofH}, and an overall target simulation error (in spectral distance) $\epsilon$ over the
simulation time $t$, the algorithm is constructed such that $\Vert U(t)-U_{\rm sim}\Vert\leq \epsilon$, where $U(t)=\e^{-iHt}$ is the exact dynamics and $U_{\rm sim}$ is its approximate circuit simulation. The overall simulation of $U(t)$ is carried out by executing $r$ consecutive simulations of $U(\Delta t)$ where $\Delta t=t/r$. In Sec.~\ref{sec:cost}, we show that an optimal choice for $r$ is  $r=t\Gamma/\ln2$ where $\Gamma=\sum_{i=1}^M \Gamma_i$  is the `off-diagonal norm' of the Hamiltonian.

To implement $U(\Delta t)$, our algorithm utilizes the LCU technique, which allows one to efficiently simulate linear combinations of simple-to-execute unitary operations. Hence, our first order of business is to approximate $U(\Delta t)$ given in Eq.~\eqref{eq:Uod} with a linear combination of simple-to-implement unitary operations with an error of at most $\epsilon/2r$, which would ensure that the overall simulation error is as desired.   

Our LCU approximation of $U(\Delta t)$ involves two steps. The first is the approximation of the divided difference exponentials appearing in the expression for the operators $A_{{\bf i}_q}$, Eq.~(\ref{eq:Aiq}), as a linear combination of phases; explicitly, we show that the following approximation holds. 
\beq
\e^{-i \Delta t [E_{z_0}\ldots,E_{z_q}]}  \approx \frac{(-i \Delta t)^q}{K^q q!} \sum_{{\bf k}_q} \e^{-i (\Delta t/K) \sum_j\alpha_j({\bf k}_q) E_{z_j}} \,,
\eeq
where  ${\bf k}_q=(k_1,\ldots,k_q)$ is a tuple of indices $k_m$, with $m=1,\ldots, q$, each ranging from $1$ to $K$ with $K$ being a positive integer whose value we determine later, and $\alpha_j({\bf k}_q)$ are real-valued coefficients that depend on ${\bf k}_q$. The above approximation is derived in  Sec.~\ref{sec:ddep}.
In Appendix~\ref{app:bound}, we obtain the bound 
\begin{align}
&\Big\vert\e^{-i \Delta t [E_{z_0}\ldots,E_{z_q}]} - \frac{(-i \Delta t)^q}{K^q q!} \sum_{{\bf k}_q} \e^{-i (\Delta t/K) \sum_j\alpha_j({\bf k}_q) E_{z_j}}\Big\vert \nn\\&\leq \frac{\Delta t^{q}}{q!}\Big(\frac{\Delta t\Delta E}{2K^2}\Big)^2 \,,
\end{align}
where $\Delta E$ is a bound on the energy differences \hbox{$|E_{z_{j+1}}-E_{z_j}|$} for all $j$, and in Sec.~\ref{sec:cost}, we show that choosing  $K={\cal O}(\mu \frac{\sqrt{r}}{\sqrt\epsilon})$, where $\mu=\Delta E/\Gamma$ ensures the desired $\epsilon$-close simulation.

The second approximation we employ involves simply truncating the PMR infinite series, Eq.~\eqref{eq:Uod}, at some maximal order $Q$. Following standard LCU analysis, we illustrate that choosing \hbox{$Q={\cal O}(\log(\Gamma t/2\epsilon)/\log\log(\Gamma t/2\epsilon))$} ensures  a simulation error of at most $\epsilon/2r$ for $U(\Delta t)$, as required. 

Once $U(\Delta t)$ is spelled out via the two approximations above as a linear combination of simple-to-implement unitary operations, we employ the LCU technique (Sec.~\ref{sec:lcu}). LCU calls for the execution of a system-ancilla unitary operator $A = -W R W^\dagger R W$, where $R$ is an ancilla reflection operation, \hbox{$R=1-2 (\ket{0}\bra{0})$}, and $W$ is defined as $W=B^\dagger U_C B$. Here, $B$ is an ancilla state preparation unitary, and $U_C$ is an ancilla-system controlled-unitary transformation, which encodes the operation of the simple unitary operations defined through $U(\Delta t)$. The application of $A$, known as Oblivious Amplitude Amplification (OAA)~\cite{Berry1}, assures that if $\Vert U(t)-\hat{U}\Vert\leq \epsilon$ then $\Vert\text{Tr}_\text{anc}(PA\ket{0}\ket{\psi}-U\ket{\psi}\bra{\psi}U^\dagger) \Vert\leq {\cal O}(\epsilon)$, where $P$ is projection of the anculla on the $\ket{0}$ state. The LCU building blocks are sketched in Fig.~\ref{fig:OAA}.

In the sections that follow, we describe in detail the simulation algorithm and rigorously derive all relevant performance bounds. 
}
\begin{figure}[h!]
\begin{flushleft}
(a)~~~\Qcircuit @C=0.7em @R=0.9em @!R{ 
&\qw&\ghost{W}&\qw&=\\
&\qw&\multigate{-1}{W}&\qw&
}
\hspace{4pt}
\Qcircuit @C=0.7em @R=0.3em @!R{ 
&  \qw&\ghost{U_C}&\qw&\qw\\
& \gate{B}&\multigate{-1}{U_C}&\gate{B^\dagger} &\qw
}\vspace{12pt}\\
(b)~~~~~~\Qcircuit @C=0.7em @R=0.3em @!R{ 
\ket{\psi(t)}  &&&\qw&  \ghost{W}&\qw&\ghost{W^\dagger}&\qw&\ghost{-W} &\qw&&&&\ket{\psi(t+\Delta t)}\\
\ket{0}   &&&\qw& \multigate{-1}{W}&\gate{R}&\multigate{-1}{W^\dagger}&\gate{R} & \multigate{-1}{-W}&\qw\\ 
}
\end{flushleft}
\caption{\ih{{\bf A circuit diagram for LCU.} (a) The circuit $W=B^\dagger U_C B$ consists of a state preparation sub-routine $B$ acting on the ancilla register and a controlled-unitary circuit $U_C$ encoding the operation of the simple unitary operations defined through $U(\Delta t)$. (b) The circuit executing a single short-time evolution step $U=\e^{-i H \Delta t}$. The top line is the system register, and the bottom line is the ancilla register.}} 
\label{fig:OAA}
\end{figure}

\subsection{Divided difference exponentials as linear combinations of phases\label{sec:ddep}} 
To begin with, we present an efficient method for calculating in superposition the divided differences appearing in the $A_{{\bf i}_q}$ operators in a manner that does not require any classical (i.e., quantum reversible) arithmetic calculations on additional ancillary registers. 

First, let us focus on the divided differences of the exponential function $ \e^{-i \tau [x_0,\ldots,x_q]}$  for a multi-set of inputs $\{x_0,\ldots,x_q\}\in\mathbb{R}^{q+1}$ and a real valued parameter $\tau$. By exploiting the Leibniz rule for divided differences~\cite{dd:67,deboor:05}, which states that 
\begin{equation}
    \label{eq:leibniz-rule}
    (f\cdot g)[x_0, \ldots, x_q] = \sum_{j=0}^q f[x_0, \ldots, x_j] g[x_j, \ldots, x_q] \,,
\end{equation}
for any two functions $f$ and $g$, we can write
\beq
 \e^{-i \tau [x_0,\ldots,x_q]} = \sum_{j=0}^{q}  \e^{-i \xi \tau [x_0,\ldots,x_j]}  \e^{-i (1-\xi) \tau [x_j,\ldots,x_q]}  \,,
 \eeq
for any $\xi \in \mathbb{C}$. A successive application of the rule yields 
\bea\label{eq:etau}
   &&\e^{-i \tau [x_0,\ldots,x_q]} =\\\nonumber
   && \sum_{\mathclap{0\leq j_1\leq \ldots \leq j_{K-1}\leq q}}\e^{-i \frac{\tau}{K} [x_0,\ldots,x_{j_1}]}  
   \e^{-i \frac{\tau}{K} [x_{j_1},\ldots,x_{j_2}]}  
\cdots\e^{-i \frac{\tau}{K} [x_{j_{K-1}},\ldots,x_{q}]}  \,,
\eea
for any integer $K \geq1$. We shall refer to $K$ as the divided differences approximation subdivision constant in what follows. For convenience, we shall choose it to be a power of two, namely, $K=2^\kappa$, for a nonnegative integer  $\kappa$ to which we will refer as the divided differences approximation depth. 

As a next step, we utilize the observation that for every finite $q$ there is small enough $\delta  \coloneqq\tau /K$, such that the divided difference  $\e^{-i \delta [x_0,\ldots,x_q]}$ is well approximated  by~\cite{kalev2024feynmanpathintegralsdiscretevariable}:
\beq\label{eq:basicApprox}
\e^{-i \delta [x_0,\ldots,x_q]} \approx \frac{(-i\delta)^q}{q!} \e^{-i \delta \frac1{q+1}\sum_{j=0}^q x_j} 
\eeq
(we discuss the quality of this approximation and its effect on the algorithm precision later on).  
We can thus use Eq.~(\ref{eq:basicApprox}) to approximate the divided differences in Eq.~(\ref{eq:etau}) to obtain 
\begin{align}\label{eq:dd_approx}
 \e^{-i \tau [x_0,\ldots,x_q]} &\approx(-i \delta)^q\sum_{\mathclap{\substack{ \\0 \leq j_1 \leq \ldots \leq j_{K-1} \leq q}}}   \frac{\e^{-i \delta (\bar{x}_{(0,j_1)}+\bar{x}_{(j_1,j_2)}+\ldots + \bar{x}_{(j_{K-1},q)})}}{j_1!(j_2-j_1)! \cdots(q-j_{K-1})!}
  \nn\\&\coloneqq \hat{\e}_K^{-i \tau [x_0,\ldots,x_q]} ,
  \end{align}
where we used the notation 
 \beq
 \bar{x}_{(j_{\ell-1},j_{\ell})}=\frac{1}{j_{\ell}-j_{\ell-1}+1}\sum_{m=j_{\ell-1}}^{j_{\ell}} x_m \,,
 \eeq
with $\ell=1,\ldots,K$ (and defining $j_0=0$). The choice for $K$ which ensures a near-optimal scaling of resources in terms of simulation error is discussed in Sec.~\ref{sec:cost}. 

Next, we write the approximation $\hat{\e}^{-i\tau[\cdots]}_K$ as a sum of simple-to-calculate phases. By a change of variables, Eq.~\eqref{eq:dd_approx}  can be rewritten as: 
\begin{align}\label{eq:e-hat}
  &\hat{\e}^{-i \tau [x_0,\ldots,x_q]}_K=(-i \delta)^q\\\nn  &\times \sum_{\mathclap{\substack{ j_1, j_2, \ldots, j_{K}\geq 0,\\ \sum_\ell j_\ell=q}}}
  \frac{   \e^{-i \delta (\bar{x}_{(0,j_1)}+\bar{x}_{(j_1,j_1+j_2)}+\ldots + \bar{x}_{(j_1+\ldots + j_{K-1},j_1+\ldots + j_{K})})} }{j_1!j_2! \cdots j_K!}\\\nn&=
\frac{(-i\delta)^q}{q!}\sum_{\sjl}
{{q}\choose{j_1,\ldots,j_K}}\prod_{\ell=1}^{K} \e^{-i\delta \bar{x}_{\ell}}\,, 
 \end{align}
where in the second equation ${{q}\choose{j_1,\ldots,j_K}}=\frac{q!}{j_1! \cdots j_K!}$ is the multinomial coefficient, and we defined
\begin{align}
\bar{x}_{\ell}&\coloneqq \bar{x}_{(j_1+\ldots + j_{\ell-1},j_1+\ldots + j_{\ell})} \\\nn&=\frac{1}{j_{\ell}+1}\sum_{m=j_1+\ldots + j_{\ell-1}}^{j_1+\ldots + j_{\ell}} x_m \,,
\end{align}
with the convention that for $\ell=1$, the sum $j_1+\ldots + j_{\ell-1} \equiv 0$. For readability, we have abbreviated $\sum_{{\substack{j_1, j_2, \ldots, j_K \geq 0, \sum_\ell j_\ell=q}}}$ by $\sum_{\sjl}$ with $\Omega_{(K,q)}$ being the set of all integer tuples ${\bf j}=(j_1,\ldots,j_K)$ that obey \hbox{$0\leq j_\ell \leq q \,, \forall \ell$,} and $\sum_{\ell=1}^K j_{\ell}=q$.

Moreover, the product  $ \prod_{\ell=1}^{K} \e^{-i\delta \bar{x}_{\ell}}$ can be recast as a product of phases, each of which is proportional to one of the original $x_j$ inputs, namely 
\bea\label{eq:ell2s}
 \prod_{\ell=1}^{K} \e^{-i\delta \bar{x}_{\ell}}
 =
 \prod_{s=0}^{q}  \e^{-i\delta \alpha_s x_s}
\eea
where the coefficients $ \alpha_s$ are given by
\begin{align}\label{eq:alpha}
\alpha_s &=\sum_{{ \ell:\,  s \in [\Sigma_{\ell-1},\Sigma_{\ell} ]}}
 \frac1{\Sigma_\ell-\Sigma_{\ell-1}+1},
\end{align}
with $\Sigma_{\ell} = j_1+ \ldots + j_\ell$.  This simplification is achieved by noticing that $x_\ell$ contributes an additive factor to $\alpha_s$ if and only if it contains $x_s$ in its average, in which case the factor is $\frac1{j_{\ell}+1} = \frac1{\Sigma_\ell-\Sigma_{\ell-1}+1}$ (otherwise, the contribution is zero). In addition, $x_{\ell}$ will contain $x_s$ if and only if $s$ is in the range $[\Sigma_{\ell-1},\Sigma_{\ell} ]$.

Furthermore, one can show that the $\alpha_s$  may be expressed in terms of the minimum $\ell$ index, which we denote $\ell_{\min}$, obeying $s \in [\Sigma_{\ell},\Sigma_{\ell+1} ]$ and the maximum $\ell$ index, denoted $\ell_{\max}$, for which $s \in [\Sigma_{\ell-1},\Sigma_{\ell} ]$, namely,
\bea\label{eq:alpha2}
\alpha_s &=&\frac1{\Sigma_{\ell_{\min}+1} - \Sigma_{\ell_{\min}}+1}+\frac1{\Sigma_{\ell_{\max}}
-\Sigma_{\ell_{\max}-1}+1}  \nonumber\\
&+& (\ell_{\max}-\ell_{\min}-2)\,.
     \eea
Using $\alpha_s$ and Eq.~\eqref{eq:ell2s} we can now write the approximation Eq.~\eqref{eq:e-hat} as a sum of phases
\begin{align}\label{eq:e-hat2}
  &\hat{\e}^{-i \tau [x_0,\ldots,x_q]}_K=
\frac{(-i\delta)^q}{q!}\sum_{{\bf k}_q}
 \e^{-i\delta \sum_s\alpha_s x_s}\,, 
\end{align}
where we replaced the sum over  $\sjl$ with a ${\bf k}_q$ multi-index. The multi-index ${\bf k}_q = (k_1,\ldots,k_q)$ is a tuple of indices $k_m$, with $m=1,\ldots, q$, each ranging from $1$ to $K$. 
The relation between the former indices $j_1,\ldots,j_K$ and the current indices $k_1,\ldots,k_q$ is such that $j_\ell$ is the number of indices in ${\bf k}_q$ whose value is $\ell$ (since there are $q$ indices, indeed $j_\ell$ can take on values in the range $[0 ,q]$). Since the ${\bf k}_q$ indices represent ordered occurrences, we remove the multinomial coefficient from the sum. We also note that $\alpha_s$ can be directly calculated from the multi-index ${\bf k}_q$ indices, since $\sum_\ell$ it is the sum total of all the index values $k_1,\ldots,k_q$ whose value is less than or equal to $\ell$. We note that the values of $\alpha_s$ coefficients are determined by ${\bf k}_q$. To make this connection explicit, we will sometimes write  $\alpha_s({\bf k}_q)$. 

Next, we describe a quantum circuit designed to approximate $U(\Delta t)$ as given in Eq.~\eqref{eq:Uod} based on the approximation derived above. 

\subsection{The LCU implementation\label{sec:lcu}}

Having obtained the approximation Eq.~\eqref{eq:e-hat2}, we insert it now into Eq.~\eqref{eq:Aiq}, identifying  $\delta=\Delta t/K$ and the inputs $x_j$  with diagonal energies $E_{z_j}$. 
Equation~\eqref{eq:Uod}  becomes
\begin{align}
U(\Delta t)\approx\hat{U}
= &\sum_{q=0}^{\infty}  \frac{\Delta t^q}{q!}    \sum_{{\bf i}_q} \Gamma_{{\bf i}_q}\frac{1}{K^q}   \sum_{{\bf k}_q}  \times \\\nn&  (-i)^q   P_{{\bf i}_q}   \left[ \sum_z \frac{d_{{\bf i}_q}}{\Gamma_{{\bf i}_q}}\e^{-i\delta\sum_{s=0}^{q}\alpha_s {E}_{z_s}}\ket{z}\bra{z}\right]  \,. 
\end{align}
At this point we assume for ease of presentation that the matrix elements $d_{{\bf i}_q}=\prod_{j=1}^q \bra{z_j}D_{i_j}\ket{z_j}$  are real-valued and independent of $z$ (i.e., that the $D_i$ matrices are proportional to the identity matrix), as this is the case for many physical systems. We will discuss the necessary adjustments to the algorithm (which do not lead to added complexity) for the case where these coefficients are $z$-dependent later on in Sec.~\ref{sec:zdep}. In this simpler case we have $D_{i}=\Gamma_i\mathds{1}$, for some $\Gamma_i\in \mathbb{R}$ and $d_{{\bf i}_q}=\prod_{j} \Gamma_{i_j}$, so $d_{{\bf i}_q}/\Gamma_{{\bf i}_q}=1$, and the evolution operator $\hat{U}$  becomes
\begin{align}\label{eq:Uod-hat}
\hat{U} = &\sum_{q=0}^{\infty}  \sum_{{\bf i}_q}  \sum_{{\bf k}_q}  \frac{\Gamma_{{\bf i}_q}\Delta t^q}{K^q q!}
  \
V_{({\bf i}_q,{\bf k}_q) }  \,,
\end{align}
where
\beq\label{eq:V}
 V_{({\bf i}_q,{\bf k}_q)} =(-i)^q   P_{{\bf i}_q}  \sum_z \e^{-i \delta \sum_{s=0}^q \alpha_s({\bf k}_q) E_{z_s}}  |z\rangle \langle z| 
 \eeq
are unitary operators. 
Truncating the infinite series in Eq.~\eqref{eq:Uod-hat} at some maximal order $Q$ and rearranging terms, we find
\begin{align}\label{eq:Uod-hat2}
 \hat{U}\approx\widetilde{U}:= \sum_{q=0}^{Q} \frac{(\Gamma \Delta t)^q}{q!} \sum_{{\bf i}_q} \frac{\Gamma_{{\bf i}_q}}{\Gamma^q} \sum_{{\bf k}_q}\frac1{K^q}   
V_{({\bf i}_q,{\bf k}_q) }  \,,
\end{align}
where we recall that $\Gamma=\sum_{i=1}^M \Gamma_i$.   The choice for $Q$ which ensures a near-optimal scaling of resources in terms of simulation error will be discussed in Sec.~\ref{sec:cost}.

Since Eq.~\eqref{eq:Uod-hat2} has the form of a linear combination of unitary operators, we can now employ the LCU technique to implement the approximate unitary.   
The technique consists of two main components: (i) the state preparation circuit $B$ on ancilla qubits and (ii) unitary circuits $U_C$ on the system state controlled by the ancilla qubits. We discuss a simple-to-implement protocol for the two in order next. 

\subsubsection{The state preparation subroutine}
The ancilla state we prepare consists of three quantum (multi-)registers:  (i) A  $Q$-qubit register, (ii) $Q$ registers with $M$ qubits each, and (iii) a third set of $Q$ registers consisting of $\kappa=\log_2K$ qubits each.
Denoting $|{\bf 1}_q\rangle =\ket{1}^{\otimes{q}}\ket{0}^{\otimes{Q-q}} $ which is the unary encoding of the expansion order $q$, 
$\ket{{\bf i}_q}=\ket{i_1}\cdots\ket{i_q}\ket{0}^{\otimes{Q-q}}$ where $\ket{i}$ is a shorthand for the $M$ qubit state $\ket{1}_i\ket{0}^{\otimes M-1}$, that is a unary encoding of $i\in[1,M]$ in which the $i$-th qubit is set to $1$ while the rest of the $M-1$ qubits are set to $0$, and $\ket{{\bf k}_q}=\ket{k_1}\cdots\ket{k_q}\ket{0}^{\otimes{Q-q}}$ -- a shorthand for $Q$ quantum registers, each of dimension $K$, the ancilla state for the LCU reads:
\begin{align}\label{eq:psi0}
|\psi_0\rangle= \frac1{\sqrt{s}} \sum_{q=0}^{Q} \sqrt{\frac{(\Gamma \Delta t)^q}{q!}} \sum_{{\bf i}_q} \sqrt{\frac{\Gamma_{{\bf i}_q}}{\Gamma^q}} \sum_{{\bf k}_q}\frac1{\sqrt{K^q}} \ket{{\bf 1}_q}\ket{{\bf i}_q}\ket{{\bf k}_q}  \,. 
\end{align}
with the normalization constant 
\begin{align}\label{eq:s}
s&=\sum_{q=0}^{Q} \frac{(\Gamma \Delta t)^q}{q!} \sum_{{\bf i}_q} \frac{\Gamma_{{\bf i}_q}}{\Gamma^q} \sum_{{\bf k}_q}\frac1{K^q} =\sum_{q=0}^Q\frac{(\Gamma\Delta t)^q}{q!}.
\end{align}

We prepare the state in three stages. In the first, we use the first $Q$-qubit register to prepare
\beq
|0\rangle^{\otimes Q} \to \frac1{\sqrt{s}} \sum_{q=0}^Q \sqrt{\frac{(\Gamma \Delta t)^q}{q!}} |{\bf 1}_q\rangle \,,
\eeq
at the cost of ${\cal O}(Q)$ controlled $R_y$ rotations~\cite{Kalev2021quantumalgorithm}.
As a next step, we use the second set of $Q$ registers to prepare
\beq\label{eqq:statePrep3}
 |{\bf 1}_q\rangle |0\rangle^{\otimes Q} \to  |{\bf 1}_q\rangle\sum_{{\bf{i}}_q} \sqrt{\frac{\Gamma_{{\bf{i}}_q}}{\Gamma^q}} |{\bf{i}}_q\rangle  \,,
\eeq
which can also be done at the cost of ${\cal O}(QM)$ controlled rotations~\cite{Kalev2021quantumalgorithm}. Note that the right-hand-side of the last equation has a tensor product structure over $q$ registers: $(\ket{1}\sum_{i=1}^M\sqrt{\Gamma_i/\Gamma}\ket{i})^{\otimes q}$. 
Similarly, the third and final stage  consists of using the last set of $Q$ $\kappa$-qubit registers to prepare
\beq
 |{\bf 1}_q\rangle |0\rangle^{\otimes Q} \to |{\bf 1}_q\rangle \frac1{\sqrt{K^q}}\sum_{{\bf k}_q}|{\bf k}_q\rangle \,, 
\eeq
which can be carried out using $Q$ controlled Hadamard gates on $\kappa$ qubits. Therefore, overall the state preparation subroutine can be completed using ${\cal O}(Q(M+\kappa))$ controlled rotations.

\subsubsection{The controlled unitaries}
\begin{figure*}[t!]
\begin{center}
\hspace{1em}\Qcircuit @C=2em @R=0.2em @!R { 
\ket{s}  &  & \ctrl{1}& \ctrl{2}& \qw&\ctrl{2}& \gate{\ll}&\qw&\\
\ket{{\bf i}_q}  & & \ctrl{3} & \qw&\qw&\qw&\qw&\qw&\\
\ket{{\bf k}_q}  & & \qw&\ctrl{1}& \qw& \ctrl{1}&\qw& \qw&\\
\ket{0} &  &  \qw &\gate{\alpha_s}& \ctrl{1}
 &\gate{-\alpha_s} &\qw&\qw&\\%
\ket{z} &  & \gate{P_{i_s}}&\qw&\gate{ie^{-i\delta\alpha_s D_0}}&\qw& \qw&\qw& \hspace{6pt}U_{(i_q,{\bf k}_q)}\ket{z} 
}
\end{center}
\caption{{\bf A circuit description of  $U_{(i_q,{\bf k}_q)}$}. $Q$ successive application of this circuit, preceded by a phase circuit $\e^{-i\delta \alpha_0D_0}$, 
make up the controlled unitary $V_{({\bf i}_q,{\bf k}_q)}$ of Eq.~\eqref{eq:selectU}. The $|s\rangle$ register consists of $Q$ qubits and is initialized with the rightmost bit set to one and the others set to zero. The \fbox{$\ll$} gate shifts the set bit one place to the left (if the leftmost bit is set, the gate sets the rightmost bit).  The $\alpha_s$ coefficients in the phase-gate $\e^{-i\delta\alpha_s D_0}$ are calculated using a circuit we denote in this figure by $\alpha_s({\bf k}_q)$. The circuit for $\alpha_s$ is described in the main text. The controlled $P_{i_s}$ circuit executes  $P_{i_1}$ then $P_{i_2}$ etc.  as per the value stored in the $|s\rangle$ register. Since the $P_i$'s are Pauli-$X$ strings, the control-$P_i$ gates consist only of simple-to-implement CNOTs.}
\label{fig:controlledU}
\end{figure*}

The second ingredient of the LCU protocol consists of designing the select-unitary operation:
\beq\label{eq:selectU}
|{\bf i}_q\rangle |{\bf k}_q\rangle |z\rangle  \to |{\bf i}_q\rangle |{\bf k}_q\rangle  V_{({\bf i}_q,{\bf k}_q)} |z\rangle \,,
\eeq
where the $V_{({\bf i}_q,{\bf k}_q)}$ are given in Eq.~\eqref{eq:V}. We now show that the select-unitary operation which executes Eq.~\eqref{eq:selectU} can be implemented using interleaved applications of CNOTs and controlled-phases operations. To that aim, we define the unitary operator:
\beq\label{eq:Uik}
U_{(i_s,{\bf k}_q)} \coloneqq (-i)   \e^{-i \delta \alpha_s ({\bf k}_q) D_0} P_{i_s}\,.
\eeq
Let us next consider the action of the ordered product
\beq
U_{(i_q,{\bf k}_q)}\cdots U_{(i_1,{\bf k}_q)} \e^{-i \delta \alpha_0 ({\bf k}_q) D_0} \coloneqq U_{({\bf i}_q,{\bf k}_q)}\e^{-i \delta \alpha_0 ({\bf k}_q) D_0}
\eeq
on a state $|z\rangle$. 
We find that 
\begin{align}
 &U_{({\bf i}_q,{\bf k}_q)}\e^{-i \delta \alpha_0 ({\bf k}_q) D_0}|z\rangle \nn\\
 &= (-i)^q   P_{{\bf i}_q}  \sum_z \e^{-i \delta \sum_{s=1}^q \alpha_s({\bf k}_q) E_{z_s}} |z\rangle=  V_{({\bf i}_q,{\bf k}_q)}|z\rangle. 
\end{align}
Hence,   $U_{({\bf i}_q,{\bf k}_q)}$  consists of an interleaved application of CNOT gates (since the $P_i$ operators are Pauli-$X$ strings) and a phase-kickback circuit that implements $\e^{-i \delta \alpha_s ({\bf k}_q) D_0}$ [up to a simple factor of $(-i)$]. A circuit for $U_{({\bf i}_q,{\bf k}_q)}$ is sketched out in Fig.~\ref{fig:controlledU}.
For simplicity, the circuit uses a $Q$-qubit ancilla that encodes $s=1, \ldots, Q$ in unary form, and a shift gate that, starting with $|0 \ldots 01 \rangle$, shifts the $1$ bit to the left with every application. By including this counter register, we are able to express $V_{({\bf i}_q,{\bf k}_q)}$ as $Q$ repeated applications of the circuit shown in Fig.~\ref{fig:controlledU}.

\ih{
In addition, since the diagonal $D_0$ is written as a linear combination of Pauli-$Z$ strings, i.e., as $D_0 = \sum J_k Z_k$, for some $J_k\in{\mathbb R}$ and where the $Z_k$ operators stand for some Pauli-$Z$ strings, the diagonal unitary $\e^{-i \delta \alpha_s ({\bf k}_q) D_0}$ is essentially a product of $\e^{-i \delta \alpha_s J_k Z_k}$ each of which is trivial to implement, once the coefficient $\alpha_s$ is provided. Each unitary operator in the above product,  $\e^{-i \delta \alpha_s J_k Z_k}$, can be further simplified as
\beq
\e^{-i \delta \alpha_s J_k Z_k}=\sum_z\e^{-i \delta \alpha_s J_k (-1)^{\sum_{l=1}^m z_l} }\ket{z}\bra{z},
\eeq
where $\ket{z}$ is the computational basis state of the $m$ qubits on which $\e^{-i \delta \alpha_s J_k Z_k}$ acts, i.e., \hbox{$z\in\{0,1\}^{m}$} and  $z_l=0,1$ is the $l$-th bit of $z$. Therefore, $\e^{-i \delta \alpha_s J_k Z_k}$ can be implemented using a single ancilla qubit and $2m$ CNOT gates~\cite{NielsenChuang}. A diagram is provided in Fig.~\ref{fig:UD0}.}
\begin{figure}[h!]
\begin{center}
\hspace{1em}\Qcircuit @C=0.75em @R=0.2em @!R { 
\ket{z_1}  & & &\ctrl{3}&\qw&\qw&\qw &\ctrl{3}&\qw\\
\vdots &  & & & & & & & & & \\
\ket{z_m}  & &  &\qw&\ctrl{1}&\qw &\ctrl{1}&\qw&\qw \\
\ket{0} &  & &\targ&\targ&\gate{\e^{-i \delta \alpha_s J_k Z_k}}&\targ &\targ &\qw
}
\end{center}
\caption{\ih{A circuit for $\e^{-i \delta \alpha_s J_k Z_k}$. The $m$ single-qubit registers control the application of  $\e^{-i \delta \alpha_s J_k Z_k}$ on the last single-qubit register.}}
\label{fig:UD0}
\end{figure}

Next, we discuss the implementation of the $\alpha_s$ calculation.

\subsubsection{Calculating the $\alpha_s$ coefficients}
First, we note that given a quantum register that stores an integer $b$, one can implement $R_z(b\theta)$ as well as $R_z(\theta/b)$ using $\log_2{b}$ calls to  a single qubit $R_z$ rotation gate~\cite{NielsenChuang}. Next, we observe that $\alpha_s$ consists of three terms, each of which is either an integer or the reciprocal thereof, cf. Eq.~\eqref{eq:alpha2}. Therefore, by calculating and storing those integers in a quantum register, one may implement $R_z(\alpha_s\theta)$ efficiently. 

Lastly, finding  $\ell_{\min}$ and $\ell_{\max}$ as per Eq.~\eqref{eq:alpha2} (each of which requiring a $\kappa$-qubit register),  may be achieved with ${\cal O}(\kappa)$ operations using reversible binary search (a blueprint for such a circuit is given in Appendix~\ref{app:binary}) provided one has access to a cost function circuit that calculates  $\Sigma_{\ell}$ for a given $\ell$ value.

A circuit for calculating  $\Sigma_{\ell}$ from $\ell$ can be constructed by implementing an integer comparison circuit~\cite{PhysRevLett.122.020502}
\beq
|k\rangle |\ell\rangle |\Sigma\rangle \to   |k\rangle |\ell\rangle |\Sigma \oplus \delta_{k \leq \ell}\rangle \,,
\eeq
i.e., a circuit that increments the third $\Sigma$ register if and only if $k \leq \ell$. Here the $|k\rangle$ and $|\ell\rangle$ registers have $\kappa$ qubits and $|\Sigma\rangle$ has $\log_2 Q$ qubits. 
Executing the above sub-routine $q$ times sequentially with the $|k_1\rangle \ldots |k_q\rangle$ functioning as the first register in each iteration, we obtain a circuit that implements 
\begin{align}
\left( |k_1\rangle |k_2\rangle \cdots |k_q\rangle\right)|\ell\rangle|0\rangle \to \left( |k_1\rangle |k_2\rangle \cdots |k_q\rangle\right)|\ell\rangle |\Sigma_{\ell} \rangle\,.
\end{align}

Implementation of a binary search circuit using $\Sigma_{\ell}$ as the cost function, allows us to construct a circuit that produces
\beq
|{\bf k}_q\rangle |0\rangle^{\otimes 4} \to 
|{\bf k}_q\rangle |\ell_{\min} \rangle 
|\ell_{\max} \rangle
|\Sigma_{\ell_{\min}}\rangle|\Sigma_{\ell_{\max}}\rangle\,.
\eeq
The last two registers can in turn be used to calculate the relevant integers that appear in Eq.~\eqref{eq:alpha2}. The above can be done in ${\cal O}(\kappa)$ operations~\cite{PhysRevLett.122.020502}.

We mention in passing that another ${\cal O}(K)$ [rather than ${\cal O}(\kappa)$] method for calculating $\alpha_s$ is available which is nonetheless simpler to implement as it does not involve the calculation of 
$\ell_{\min}$ and $\ell_{\max}$. For this one, we write
\beq
\e^{-i\delta \alpha_s E_{z_s}} = \prod_{\ell=1}^{K} \e^{-i \delta
 \frac{ b_{s \ell} E_{z_s}}{\Sigma_\ell-\Sigma_{\ell-1}+1}}\,,
 \eeq
 where $b_{s \ell}$ denotes a bit  that is set to one if $s \in [\Sigma_{\ell-1},\Sigma_{\ell} ]$ and is zero otherwise. This requires a circuit $|s\rangle |\ell\rangle |0 \rangle \to |s\rangle |\ell\rangle |b_{s \ell} \rangle$ implementing an integer comparison circuit which checks for the conditions $s \geq \Sigma_{\ell-1}$ and $s \leq \Sigma_{\ell}$.
 
\subsection{Algorithm cost\label{sec:cost}}
In the previous section, we worked out the circuit $\widetilde{U}$ which approximates the short-time evolution $U(\Delta t)$. In this section we provide the resource scaling analysis that ensures that $\widetilde{U}$ is $\epsilon/r$-close to $U(\Delta t)$ in spectral distance, $\Vert\widetilde{U}-U(\Delta t)\Vert\leq \epsilon/r$, where $\epsilon$ is the overall error over the simulation time $t$, and  $r=t/\Delta t$. From the subadditivity property of the spectral norm, it is ensured that with $r$ repetitions of $\widetilde{U}$, the overall simulation is $\epsilon$-close to the exact dynamics induced by the exact $\e^{-iHt}$.

Our algorithm involves two basic approximations to the exact dynamics, $U(\Delta t)\to \hat{U}\to\widetilde{U}$, determined by the divided differences approximation constant $K$ and the truncation order $Q$, respectively.  Therefore, to ensure that $\Vert\widetilde{U}-U(\Delta t)\Vert\leq \epsilon/r$ we require that both $\Vert\hat{U}-U(\Delta t)\Vert$ and $\Vert\widetilde{U}-\hat{U}\Vert$  are at most $\epsilon/2r$.

First, we note that the LCU formalism~\cite{Berry1} dictates that the ancilla state preparation normalization factor $s$ should be such that $|s-2|\leq \epsilon/2r$  to ensure that $\Vert\widetilde{U}-\hat{U}\Vert\leq \epsilon/2r$.  According to Eq.~\eqref{eq:s}, fixing $\Delta t=\ln2/\Gamma$ (equivalently, $r=t\Gamma/\ln2$) and $Q={\cal O}(\log(\Gamma t/2\epsilon)/\log\log(\Gamma t/2\epsilon))$, in our algorithm ensures $|s-2|\leq \epsilon/2r$, as required.

Second, based on Eq.~\eqref{eq:basicApprox}, we show in Appendix~\ref{app:bound} that
\beq\label{eq:bigApprox}
\Big\vert\e^{-i \Delta t [E_{z_0}\ldots,E_{z_q}]} - \hat{\e}^{-i \Delta t [E_{z_0}\ldots,E_{z_q}]}_K\Big\vert \leq \frac{\Delta t^{q}}{q!}\Big(\frac{\Delta t\Delta E}{2K^2}\Big)^2 \,,
\eeq
where $\Delta E$ is a bound on the energy differences \hbox{$|E_{z_{j+1}}-E_{z_j}|$} for all $j$, i.e., on two consecutive diagonal energies. 
From the above, it follows that.   
\begin{align}
\Vert\hat{U}-U(\Delta t)\Vert&\leq\sum_{q,{\bf i}_q} \Gamma_{{\bf i}_q}  \Big| A_{{\bf i}_q} -\hat{A}_{{\bf i}_q} \Big|\\\nn
&=\sum_{q,{\bf i}_q} \Gamma_{{\bf i}_q} \frac{\Delta t^q}{q!} \left( \frac{\Delta t \Delta E}{2 K} \right)^2=\frac1{2}
\left( \frac{\Delta t \Delta E}{K} \right)^2.
\end{align}
Denoting by $\mu$ the ratio $\Delta E/\Gamma$,  i.e., the ratio between the off-diagonal norm which we expect to scale with system size and the energy difference between two basis states separated by a single hop, which is not, we find that 
choosing $K={\cal O}(\mu \frac{\sqrt{r}}{\sqrt\epsilon})$, equivalently 
\hbox{$\kappa={\cal O}(\log \mu \sqrt{t \Gamma /\epsilon})$},  ensures that $\Vert\hat{U}-U(\Delta t)\Vert\leq{\epsilon/(2r)}$. 

Finally, with the above choices for $Q$ and $\kappa$, we recall that the number of ancilla qubits needed and the number of gates of a single LCU execution are  ${\cal O}(Q(M + \kappa))$, as discussed above. Since both $Q$ and $\kappa$ scale as $\log t \Gamma / \epsilon$, we find that the algorithm does indeed have near-optimal dependence of precision. 

\subsection{The case of $z$-dependent $d_{{\bf i}_q}$\label{sec:zdep}}
\ih{
In the previous section, we assumed for simplicity that the product of off-diagonal matrix elements $d_{{\bf i}_q}/\Gamma_{{{\bf i}_q}}$ simplifies to $d_{{\bf i}_q}/\Gamma_{{{\bf i}_q}}=1$ whereas in the general case all that is guaranteed is that $|d_{{\bf i}_q}/\Gamma_{{{\bf i}_q}}| \leq 1$ which follows from the fact each element in the product is similarly bounded: $|d_{i_j}/\Gamma_{i_j}|\leq1\quad \forall j$. The $z$-dependent ratio  $d_{i_j}/\Gamma_{i_j}$ may always be expressed as the average of two ($z$-dependent) phases: 
\beq
\frac{d_{i_j}}{\Gamma_{i_j}}=\frac1{2}\Big(\e^{i (\theta_{i_j}+\phi_{i_j})} + \e^{i(\theta_{i_j}-\phi_{i_j})}\Big).
\eeq
From the above observation, it follows that one can simply replace the unitary $U_{(i_s,{\bf k}_q)}$ in Eq.~\eqref{eq:Uik} with 
\beq
U_{(i_s,{\bf k}_q,\pm)} \coloneqq  -i \Phi_s^{(\pm)} \e^{-i \delta \alpha_s ({\bf k}_q) D_0} P_{i_s},
\eeq
where $\Phi_s^{
(\pm)} = \sum_z \e^{i (\theta_{i_s} \pm \phi_{i_s})} |z\rangle \langle z|$ are diagonal phase unitaries.
The simulation circuit in this case can be implemented in much the same manner as before, where now the factor  $\e^{i (\theta_{i_s}+\pm \phi_{i_s})}$ can be encoded directly as a phase if the off-diagonal matrix elements are given in polar coordinates (and otherwise using a phase-kickback circuit 
using quantum registers~\cite{Kalev2021quantumalgorithm}).}
Additionally, the LCU state preparation should be modified so as to include $Q$ additional qubits each prepared in the $\ket{+}$ state alongside the change $\Gamma \Delta t \to \Gamma \Delta t /2$ in the first stage of the state preparation routine to account for the $q$ extra factors of $1/2$. These modifications do not alter the overall complexity of the algorithm.

\section{Examples\label{sec:results}}

To demonstrate the scaling advantages that the PMR method provides over existing schemes, we analyze in this section the asymptotic algorithmic cost of the dynamical simulation of two prominent physical models: the Rydberg atom Hamiltonian and dipolar fermions in optical lattices.

\subsection{The Rydberg atom Hamiltonian} 

The Rydberg atom Hamiltonian~\cite{lukin2001dipole,saffman2010quantum,weimer2010long,bernien2017probing,dudin2012controlled} 
describes the dynamics of a system of atoms excited to high-energy Rydberg states, incorporating both coherent driving by a laser and long-range interactions between atoms, leading to the Rydberg blockade phenomenon, where one atom in the Rydberg state suppresses excitation of nearby atoms. 
An $N$-atom Rydberg Hamiltonian is typically written as:
\bea\label{eq:ryd}
H_{\textrm{Ryd}} = \frac1{2} \sum_{i=1}^N  \left( \Omega_{i} X_i -\delta  Z_i\right) + \sum_{i < j} \frac{C_6}{|\mathbf{r}_i - \mathbf{r}_j|^6} n_i n_j, \nonumber\\
\eea
where the $\Omega_i$ are the Rabi frequencies, which describe the strength of the lasers driving the transition between the ground and Rydberg states of the atoms,  $\frac{\delta}{2} Z_i$ represent the detunings of the driving laser and the number operator $n_i$ is defined as $n_i=\frac1{2}\left( \mathds{1}+Z_i\right)$, projecting the state of the atom at site $i$ onto the Rydberg state. The terms $C_6/|\mathbf{r}_i - \mathbf{r}_j| n_i n_j$ correspond to the van der Waals interactions between two atoms in their Rydberg states.

In PMR form, the Rydberg atom Hamiltonian reads $H_{\textrm{Ryd}} = D_0 + \sum_i D_i P_i$ with the diagonal component being
\beq
D_0 = -\frac{\delta}{2} \sum_{i=1}^N  Z_i+ \sum_{i < j} \frac{C_6}{|\mathbf{r}_i - \mathbf{r}_j|^6} n_i n_j \,,
\eeq
and where $P_i=X_i$ and $D_i= \Omega_i \cdot \mathds{1}$ (for $i>0$). From this decomposition, one immediately finds that the off-diagonal norm for this Hamiltonian is \hbox{$\Gamma = \sum_i \Omega_i= N\Omega$} (where $\Omega$ is the average frequency)  and the number of terms in the PMR decomposition is $M=N$, the number of atoms. Recalling that the number of repetitions $r$, and hence the total cost of the PMR-based simulation algorithm, is linear in $\Gamma t=t N \Omega$ and that the qubit and gates costs, per repetition, are linear in $M$, we find that the total algorithmic cost of a dynamical simulation using the PMR method is quadratic in $N$. Importantly, the cost of the proposed algorithm is independent of the system characteristics, $\delta$ and $C_6$, which can vary from one Rydberg system to another. 

In stark contrast, when considering state-of-the-art simulation algorithms, which normally invoke a Pauli-string decomposition of the Hamiltonian, the total term count is $M'={\cal O}(N^2)$ in the decomposition of the Hamiltonian. In addition, total Hamiltonian norm, which determines the simulation cost of such algorithm is $\Gamma' = N(\Omega+\frac{\delta}{2}) + \sum_{i<j}\frac{C_6}{|\mathbf{r}_i - \mathbf{r}_j|^6}$. Hence, $\Gamma'$ is generally ${\cal O}(N)$ times larger than the relevant norm of the PMR method. Thus, compared to our algorithm, whose total cost scales as $M\Gamma\sim{\cal O}(N^2)$, the cost of comparable algorithms scales as $M'\Gamma'\sim{\cal O}(N^4)$. Moreover, the exact cost of the latter schemes depends on the $\delta$ and $C_6$ coefficients as well as on the distances between the atoms -- a dependence that is absent from the PMR cost analysis. 
    
\subsection{Dipolar fermions in optical lattices}

The second system we analyze is that of dipolar fermions in optical lattices~\cite{Lahaye2009,Baranov2012,Kadau2016,Buchler2007,Choi2014,Ferlaino2015}. These are systems of fermionic particles with intrinsic dipole moments, such as polar molecules or magnetic atoms, confined in periodic potentials created by interfering laser beams. Dipolar fermions in optical lattices provide an ideal platform for studying quantum many-body physics with long-range and anisotropic interactions, enabling the exploration of exotic quantum phases and phenomena, and are used for modeling strongly correlated systems, high-temperature superconductivity, and exotic quantum materials. 

The system is modeled using an extended Fermi-Hubbard Hamiltonian:
\bea\label{eq:dfol}
    H_{\textrm{dpol}} &=& -t_{\textrm h} \sum_{\langle i, j \rangle,\sigma} c_{i\sigma}^\dagger c_{j\sigma} 
    + U \sum_i n_{i\uparrow} n_{i\downarrow}    \nonumber\\ &+& \frac{1}{2} \sum_{i \neq j} \frac{C_\text{dd}}{|i - j|^3} 
    \left[ 1 - 3 \frac{(d \cdot r_{ij})^2}{|r_{ij}|^2} \right] n_i n_j ,
\eea
where $t_{\textrm h}$ is the hopping amplitude between nearest-neighbor sites $\langle i, j \rangle$, $C_\text{dd}$ is the dipole-dipole interaction strength,  $n_i = c_{i\uparrow}^\dagger c_{i\uparrow}+c_{i\downarrow}^\dagger c_{i\downarrow}$ is the number operator, $U$ represents the on-site interaction, and the angular term $\left[ 1 - 3 \frac{(d \cdot r_{ij})^2}{|r_{ij}|^2} \right]$ describes the anisotropy of dipole-dipole interactions. 

To simulate the dynamics of the model on a quantum computer, one customarily maps the model to an equivalent qubit model. This can be done in a number of different ways~\cite{jwt,BK02,Verstraete_2005,DK20} the most common of which is  the Jordan-Wigner transformation (JWT)~\cite{jwt} which maps the second-quantized operator  $c_{j \sigma}$ to an operator on $j$ qubits according to
\beq
c_{j \sigma} \to\left( \prod_{k=1}^{j-1} Z_{k \sigma} \right)\frac{X_{j \sigma}-i Y_{j \sigma}}{2}
\eeq
so that \hbox{$c^{\dagger}_{j \sigma} c_{j \sigma} = (\mathds{1}+Z_{j \sigma})/2$}.
Applying the transformation to the Hamiltonian, Eq.~\eqref{eq:dfol}, we arrive at:
\beq
H_{\textrm{dpol}} = D_0 +\sum_{\langle i j \rangle\sigma} D_{ij\sigma} X_{i\sigma} X_{j\sigma} \,,
\eeq
where we have identified 
\begin{align}
D_0 &= \frac{U}{4} \sum_{j=1}^N (\mathds{1}+Z_{j \uparrow})(\mathds{1}+Z_{j \downarrow}) 
\\\nonumber 
&+\sum_{i \neq j} \gamma_{ij}(\mathds{1}+\frac{1}{2}Z_{i \uparrow}+\frac1{2}Z_{i \downarrow})(\mathds{1}+\frac1{2}Z_{j \uparrow}+\frac1{2}Z_{j \downarrow}) \,.
\end{align}
Here, we have defined $ \gamma_{ij}=\frac{1}{2}  \frac{C_\text{dd}}{|i - j|^3} 
    \left[ 1 - 3 \frac{(d \cdot r_{ij})^2}{|r_{ij}|^2} \right]$
and 
\begin{align}
D_{ij\sigma} = -\frac1{4} t_{\textrm h} \prod_{k=i}^{j-1} Z_{k\sigma}\,.
\end{align}
The product structure of $D_{ij\sigma}$ implies that their max-norm is simply given by $\frac1{4}t_{\textrm h}$ for all $i,j,\sigma$ yielding an off-diagonal norm of $\Gamma=\frac1{4}t_{\textrm h} N d$, where $N$ is the number of sites in the optical lattice and $d$ is its dimension. Furthermore, the number of off-diagonal terms is $M=N d$. Therefore, the total simulation cost of our proposed algorithm is ${\cal O}(t N^2d^2)$. For comparison, in existing methods, the Hamiltonian is decomposed in terms of Pauli strings, the number of terms in the Hamiltonian is $M'={\cal O}(N+Nd+N^2)$, i.e., scales quadratically rather than linearly with the number of sites. In addition, the number of short-time evolutions (number of repetitions) is proportional to $\frac1{4}t_{\textrm h} N d+ N(N-1)/2 \gamma  + N U$ where $\gamma$ is the averaged dipole-dipole interaction strength in the system. We therefore find that, in contrast to PMR, the total simulation cost of existing algorithms is of the order $N^2d^2+N^3d+N^4$ -- a scaling that could easily become cost-prohibitive in near-term devices. 

\section{Summary and conclusions\label{sec:summary}}
 
We devised a simple-to-implement quantum algorithm designed to simulate the dynamics of general Hamiltonians on a quantum computer. While straightforward to execute, we have shown that the proposed algorithm retains near-optimal dependence on the target precision.

Our algorithm has numerous properties that we argue make it attractive for implementation on resource-limited near-future quantum computing devices, which are characterized by being small and noisy, and on which one is not afforded the luxury of fault tolerance and error correction schemes. These are: (i) Neither the gate cost nor the qubit cost of the algorithm depends on the norm of the diagonal component of the Hamiltonian. While for most simulation algorithms the number of repetitions of the short-time evolution unitary grows linearly with the diagonal norm, in the present algorithm, the dependence on the diagonal part $D_0$ enters only via trivial phases. (ii) In addition, the present algorithm offers a compact LCU decomposition of the Hamiltonian in which the terms in the Hamiltonian are grouped according to which bits they flip. This is the PMR decomposition, which is considerably more compact than the customary breakup of $H$ to a linear combination of Pauli strings in the general case~\cite{Kalev2021quantumalgorithm}. As such, the number of ancilla qubits required for the LCU state preparation subroutine is expected to be, in general, less costly by orders of magnitude when compared to standard decompositions. (iii) Last, we note again the simplicity of our proposed algorithm, which prescribes an implementation of only simple, and for the most part, straightforward, sub-routines consisting only of CNOTs and controlled phase operations. Unlike existing techniques offering near-optimal scaling with error, which require compilation and ``classical'' calculations and other such hidden costs, such as complicated phases or trigonometric functions, the algorithm proposed here does not include any such costs. 
The above property is especially important in the NISQ era, where quantum computing platforms are small and noisy, and where any unnecessary qubit or gate that is added to the circuit may be detrimental to performance. We have demonstrated the above advantages by analyzing two physical models, the Rydberg atom Hamiltonian and dipolar molecules in optical lattices, illustrating the algorithm's scaling advantages over existing schemes. 

Furthermore, the PMR method presented above also extends naturally to the case of time-dependent Hamiltonians with the main modification being that in the time-dependent case, the divided-difference inputs no longer consist only of diagonal energies but must be augmented with the frequencies of the time dependence~\cite{PRXQuantum.2.030342}, namely, 
\beq
E_{z_j} \to E_{z_j} + \sum_{\ell = j+1}^q \lambda_{(\ell,z_j)} 
\eeq
where the $\lambda_{(\ell,z_j)} $ are (in general complex-valued) frequencies of the now time-dependent diagonal operators $D_i$ in the PMR decomposition of the Hamiltonian. As this is the only difference between the time-independent and time-dependent cases, the divided-difference approximation introduced here similarly applies to time-dependent simulations as well. A major advantage of the PMR formulation in the time-dependent case is that the cost of the present algorithm is linear in the evolution time and does not depend on frequencies (see Ref.~\cite{PRXQuantum.2.030342} for more details). This property translates to potentially critical savings in gate and qubit costs as well.

In light of the above, we hope that the algorithm proposed here will prove to be a useful tool in the coming years, where near-term quantum computing devices become more widely available and allow their users to generate credible simulation results for scientifically relevant problems. In a future study, we hope to report on additional performance comparisons of this algorithm against existing schemes when tasked with simulating a scientifically meaningful model and executed on a NISQ device. An apples-to-apples comparison against existing algorithms on specific applications will highlight the advantages of our approach.

\begin{acknowledgments}
We thank Arman Babakhani and Lev Barash for useful suggestions. 
IH acknowledges support by the Office of Advanced Scientific Computing Research of
the U.S. Department of Energy under Contract No DE-SC0024389.
In addition, this research was developed with funding from the Defense Advanced Research Projects Agency (DARPA) under Contract No. HR00112330014. The views, opinions, and/or findings expressed are those of the authors and should not be interpreted as representing the official views or policies of the Department of Defense or the U.S. Government.

\end{acknowledgments}
\bibliography{refs}

\begin{widetext}

\begin{appendix}

\section{Notes on divided differences\label{app:dd}}

We provide below a brief summary of the concept of divided differences, which is a recursive division process. This method is typically encountered when calculating the coefficients in the interpolation polynomial in the Newton form.

The divided differences~\cite{dd:67,deboor:05} of a function $f(\cdot)$ are defined as
\beq\label{eq:divideddifference2}
f[x_0,\ldots,x_q] \equiv \sum_{j=0}^{q} \frac{f(x_j)}{\prod_{k \neq j}(x_j-x_k)}
\eeq
with respect to the list of real-valued input variables $[x_0,\ldots,x_q]$. The above expression is ill-defined if some of the inputs have repeated values, in which case one must resort to the use of limits. For instance, in the case where $x_0=x_1=\ldots=x_q=x$, the definition of divided differences reduces to: 
\beq
f[x_0,\ldots,x_q] = \frac{f^{(q)}(x)}{q!} \,,
\eeq 
where $f^{(n)}(\cdot)$ stands for the $n$-th derivative of $f(\cdot)$.
Divided differences can alternatively be defined via the recursion relations
\bea\label{eq:ddr}
f[x_i,\ldots,x_{i+j}] = \frac{f[x_{i+1},\ldots , x_{i+j}] - f[x_i,\ldots , x_{i+j-1}]}{x_{i+j}-x_i} \,,
\eea 
with $i\in\{0,\ldots,q-j\},\ j\in\{1,\ldots,q\}$ and the initial conditions
\beq\label{eq:divideddifference3}
f[x_i] = f(x_{i}), \qquad i \in \{ 0,\ldots,q \}  \quad \forall i \,.
\eeq
A function of divided differences can be defined in terms of its Taylor expansion
\beq
f[x_0,\ldots,x_q] = \sum_{n=0}^\infty \frac{f^{(n)}(0)}{n!} [x_0,\ldots,x_q]^n \ .
\eeq
Moreover, it is easy to verify that
\beq  \label{eq:ts}
[x_0,\ldots,x_q]^{q+n} = \Bigg\{ 
\begin{tabular}{ l c l }
   $0$ & \phantom{$0$} & $n<0$ \\
  $1$ & \phantom{$0$} &  $n=0$ \\
  $\sum_{\sum k_j = n} \prod _{j=0}^{q} x_j^{k_j}$& \phantom{$0$} &  $n>0$ \\
\end{tabular}
 \,.
\eeq
One may therefore write:
\beq
f[x_0,\ldots,x_q] = \sum_{n=0}^\infty \frac{f^{(n)}(0)}{n!} [x_0,\ldots,x_q]^n =
\sum_{n=q}^\infty \frac{f^{(n)}(0)}{n!} [x_0,\ldots,x_q]^n =
\sum_{m=0}^\infty \frac{f^{(q+m)}(0)}{(q+m)!} [x_0,\ldots,x_q]^{q+m}.
\eeq
The above expression can be further simplified to
\beq
f[x_0,\ldots,x_q] =
\sum_{\{ k_i\}=(0,\ldots,0)}^{(\infty,\ldots,\infty)}
\frac{f^{(q+\sum k_i)}(0)}{(q+\sum k_i)!} \prod _{j=0}^{q} x_j^{k_j} \,.
\eeq
\ih{In the main text, the function $f(\cdot)$ that arises in the unitary time evolution expansion is 
$f(\cdot)= \e^{-i \tau (\cdot)}$ where the real-valued inputs are diagonal energies (diagonal entries of the Hamiltonian). Therefore expressions of the form $\e^{-i \tau  [E_0,\ldots,E_q]}$ which in the case on non-repeating inputs can be defined as 
\beq
e^{-i \tau [E_0,\ldots,E_q]}=\sum_{j=0}^{q} \frac{e^{-i \tau E_j}}{\prod_{k \neq j}(E_j-E_k)} \,,
\eeq
are used throughout.}

\section{Derivation of the divided-difference approximation bound\label{app:bound}}

Here, we bound the absolute value of the difference between the divided difference exponential
$\e^{-i \Delta t [E_0,\ldots,E_q]}$ and its approximation
\beq\label{eq:b1}
\hat{\e}_K^{-i \Delta t [E_{z_0},\ldots,E_{z_q}]} \coloneqq
\left(\frac{-i\Delta t}{K}\right)^q\sum_{\mathclap{\substack{ \\0 \leq j_1 \leq \ldots \leq j_{K-1} \leq q}}}   \frac{\e^{-i \frac{\Delta t}{K} (\bar{E}_{(0,j_1)}+\bar{E}_{(j_1,j_2)}+\ldots + \bar{E}_{(j_{K-1},q)})}}{j_1!(j_2-j_1)! \cdots(q-j_{K-1})!} \,,
\eeq 
where $\bar{E}_{(j_{\ell},j_{\ell+1})} = \frac{E_{z_{j_{\ell}}}+ \cdots +E_{z_{j_{\ell+1}}}}{j_{\ell+1}-j_\ell+1}$. 
For what follows, we shall assume that $|E_{z_{j+1}}-E_{z_{j}}| \leq \Delta E$ for some $\Delta E$ which is of order unity, i.e., we shall assume it is an ${\cal O}(1)$ quantity that does not scale with system size, evolution time or Hamiltonian norm. This will be the case for all physical Hamiltonians, as the basis states $|z_{j+1}\rangle$ and $|z_{j}\rangle$  differ for those by a single $k$-local permutation operation. Thus, for any the physical Hamiltonian the change $\Delta E$ corresponds to a $k$-local change in the basis state energy. 
 
Next, let us use the fact~\cite{kalev2024feynmanpathintegralsdiscretevariable} that the divided difference approximation is worst when the standard deviation of its inputs is maximal, namely 
\beq
\left\vert
 \frac{q! \e^{-i \Delta t [E_{z_0},\ldots,E_{z_q}]}}{(-i \Delta t )^q   \e^{-i \Delta t (E_{(0,q)})} } - 1 \right\vert \leq  \frac{\sigma^2 \Delta t^2}{2(q+2)} \,,
 \eeq
 where $\sigma^2$ is the variance of the inputs $\{E_{z_0},\ldots,E_{z_q}\}$. Owing to this observation, we find that the bound 
 \beq\label{eq:bound_DeltaE}
\Big\vert\e^{-i \Delta t [E_{z_0},E_{z_1}\ldots,E_{z_q}]}-\hat{\e}_K^{-i \Delta t [E_{z_0},\ldots,E_{z_q}]}\Big\vert
=\Big\vert\e^{-i \Delta t [0,E_{z_1}-E_{z_0}, \ldots, E_{z_q}-E_{z_0}]}-\hat{\e}_K^{-i \Delta t [0,E_{z_1}-E_{z_0}, \ldots, E_{z_q}-E_{z_0}]}\Big\vert
\eeq
is maximized for the inputs $E_{z_j} = j \Delta E$ (i.e., this choice maximizes the variance of the inputs). Equation~\eqref{eq:bound_DeltaE} can thus be bounded by the quantity
 \beq
 \Big\vert\e^{-i \Delta t [0 ,\Delta E, \ldots, q \Delta E]}-\hat{\e}_K^{-i \Delta t [0 ,\Delta E, \ldots, q \Delta E]}\Big\vert \,,
\eeq
a quantity that we can calculate exactly, as both the divided difference and its approximation can be evaluated directly. 
A calculation of $\e^{-i \Delta t [0 ,\Delta E, \ldots, q \Delta E]}$ reveals:
\beq
\e^{-i \Delta t [0 ,\Delta E, \ldots, q \Delta E]} = \frac{ \left( \frac{-2 i \e^{-i  \Delta t \Delta E /2}}{\Delta E} \sin \frac{\Delta t \Delta E}{2} \right)^q}{q!} \,, 
\eeq
whereas for $\hat{\e}_K^{-i \Delta t [E_{z_0},\ldots,E_{z_q}]}$, after plugging in the inputs and simplifying, we obtain
\beq
\hat{\e}_K^{-i \Delta t [0,\Delta E, \ldots,q \Delta E]} = 
\frac{ \left( \frac{-2 i \Delta t \e^{-i  \Delta t \Delta E /2}}{2 K \sin \frac{\Delta t \Delta E}{2K}} \sin \frac{\Delta t \Delta E}{2} \right)^q}{q!} \,,
\eeq
which gives for their ratio
\beq
\frac{\e^{-i \Delta t [0 ,\Delta E, \ldots, q \Delta E]} }{\hat{\e}_K^{-i \Delta t [E_{z_0},\ldots,E_{z_q}]} } = \left( \frac{\sin\frac{\Delta t \Delta E}{2K}}{\frac{\Delta t \Delta E}{2K}} \right)^q \,.
\eeq
Moreover, 
\beq
\left\vert \hat{\e}_K^{-i \Delta t [E_{z_0},\ldots,E_{z_q}]}\right\vert  = \frac{(\Delta t/K)^q}{q!} \left(\frac{\sin\frac{\Delta t \Delta E}{2}}{\sin\frac{\Delta t \Delta E}{2 K}} \right)^{q}
\leq
\frac{(\Delta t/K)^q}{q!} \left(\frac{\sin\frac{\Delta t \Delta E}{2}}{\frac{\Delta t \Delta E}{2 K}} \right)^{q} = \frac{1}{q!} \left(\frac{\sin\frac{\Delta t \Delta E}{2}}{\frac{\Delta E}{2}} \right)^{q}
\leq  \frac{\Delta t^q}{q!}  \,,
\eeq
a bound that can also be read off Eq.~(\ref{eq:b1}). 
Putting it all together, we find that
\bea
\Big\vert\e^{-i \Delta t [E_{z_0},E_{z_1}\ldots,E_{z_q}]}-\hat{\e}_K^{-i \Delta t [E_{z_0},\ldots,E_{z_q}]}\Big\vert 
&\leq&
\left\vert \hat{\e}_K^{-i \Delta t [E_{z_0},\ldots,E_{z_q}]} \right\vert
\left\vert\frac{\e^{-i \Delta t [0 ,\Delta E, \ldots, q \Delta E]} }{\hat{\e}_K^{-i \Delta t [E_{z_0},\ldots,E_{z_q}]} }-1\right\vert
\\\nonumber 
&\leq&\frac{\Delta t^q}{q!}\left(1- \left(\frac{\sin\frac{\Delta t \Delta E}{2K}}{\frac{\Delta t \Delta E}{2K}} \right)^{q}\right)
 \leq \frac{\Delta t^q}{q!}   \left( \frac{\Delta t \Delta E}{2 K} \right)^2\,. 
\eea

\section{A quantum circuit for reversible binary search}\label{app:binary}
Given an oracle $|x\rangle |0\rangle \to |x\rangle |f(x)\rangle$ where $f(x)$ is monotonically non-decreasing in $x$ and $x=0 \ldots K -1$ (here K$=2^\kappa$), the task of a reversible binary search circuit is to find the smallest index $x_*\in[0,K-1]$ such that $s \geq f(x_*)$ for a given input $s$, namely, a circuit that yields $|s\rangle |0\rangle \to |s\rangle |x_*\rangle$. Here, the second register has $\kappa$ qubits. 

To construct the desired circuit, we utilize an integer comparison oracle $C |s\rangle |f\rangle |0\rangle =|s\rangle |f\rangle |b_{sf}\rangle$ where $b_{sf}=1$ iff $s \geq f$ (see Ref.~\cite{PhysRevLett.122.020502}). We implement $|s\rangle |0\rangle \to |s\rangle |x_*\rangle$ by $\kappa$ calls to $C$, such that each call will fix a single bit of $x_*$ going from left (most significant) to right (least significant). 

Starting with the leftmost bit of $x_*$, which we call $x_{\kappa}$, and moving to the right, we have $C |s\rangle |f(K/2)\rangle |0\rangle\ket{0}^{\otimes \kappa-1}=|s\rangle |f(K/2)\rangle |x_{\kappa}\rangle\ket{0}^{\otimes \kappa-1}$. We note that the input to $f$, namely $K/2$, is the value of the output register with the leftmost bit set to $1$. The next bit is similarly set by $C |s\rangle |f(x_{\kappa} K/2+ K/4)\rangle |x_\kappa \rangle\ket{0}\ket{0}^{\otimes \kappa-2}=|s\rangle |f(x_{\kappa} K/2+ K/4)\rangle |x_\kappa \rangle\ket{x_{\kappa-1}}\ket{0}^{\otimes \kappa-2}$, noting that now the input to $f$ is the value of the output register with the second leftmost bit set to 1. This sets $x_{\kappa-1}$. We continue similarly to set the rest of the $x_*$ bits.  After $\kappa$ iterations of we find the output register is the desired index $x_*$. 

\end{appendix}

\end{widetext}

\end{document}